\documentclass[conference,compsoc]{IEEEtran}
  \usepackage{natbib}
  \setcitestyle{numbers,open={[},close={]}}
\ifCLASSINFOpdf
  \usepackage{graphicx}
  \graphicspath{{Fig/}}
  \usepackage{placeins}
\else
\fi

\usepackage{ragged2e}

\begin{document}
\twocolumn[
\begin{@twocolumnfalse}
\title{Documenting Bioinformatics Software Via Reverse Engineering}

\author{\IEEEauthorblockN{Vinicius Soares Silva Marques}
\IEEEauthorblockA{Federal University of Uberlandia (UFU)\\
Institute of Biotechnology (IBTEC)\\
Patos de Minas, Minas Gerais, Brazil\\
Email: marques.vinicius@gmail.com}
\and
\IEEEauthorblockN{Laurence Rodrigues do Amaral}
\IEEEauthorblockA{Federal University of Uberlandia (UFU)\\
Faculty of Computation (FACOM)\\
Patos de Minas, Minas Gerais, Brazil\\
Email: laurence@ufu.br}
}


% compsoc conferences compared to traditional conferences), use this
%Telephone: (800) 555--1212, Fax: (888) 555--1212}


\maketitle

\justifying{
\begin{abstract}
Documentation is one of the most neglected activities in Software Engineering, although it is an important method of assuring quality and understanding. Bioinformatics software is generally written by researchers from fields other than Computer Science who usually do not provide documentation. Documenting bioinformatics software may ease its adoption in multidisciplinary teams and expand its impact on the community. In this paper, we highlight how one can document software that is already finished, using reverse engineering and thinking of the end-user.
\end{abstract}
}
\Centering

\end{@twocolumnfalse}]
% \fi
%\IEEEpeerreviewmaketitle

\section{Introduction}

Bioinformatics software are computational systems built for Biology and Biotechnology applications \cite{jawdat2006era}. They are usually written and used by multidisciplinary teams, which often include researchers from fields as diverse as Biology, Biochemistry, Genetics, Pharmacy and Medicine \cite{kumar2017role}. While some of these teams do have Computing researchers, some are comprised of other fields scientists who learned to write software. They may not be aware of the best Software Engineering practices, and usually don’t provide the documentation alongside their software. Thus, even though the software itself may be efficient and effective, it may be difficult for other researchers outside the original team to use it.

Software documentation is one of the most important tasks in Software Engineering \cite{kipyegen2013importance}, though it is one of the most neglected \cite{parnas2011precise}. Time and cost restraints often lead to poor or incomplete documentation \cite{lethbridge2003software}. But, considering both maintaining and end-user points of view, documentation provides valuable information, from an overview of the system to technical detailing of how the software works, being useful for developers, maintainers, end-user, and the community \cite{parnas2011precise}.

The application of Software Engineering to Bioinformatics has been also neglected, partly because of a creed that the use of software in scientific contexts is different than in commercial situations \cite{lawlor2015engineering}. This seems to come from a misunderstanding of concepts from computer programming and Software Engineering itself, which leads researchers from other fields who know how to code to develop software without the best practices in mind, slowing discoveries and making them less reliable than they could be. Bioinformatics researchers usually are self-taught programmers \cite{verma2013lack} that know the scientific method considers documenting all processes a good practice, but due to the lack of formal training in Software Engineering don’t document the software they develop.

Documentation is important not only to guide the development process but also to conduct the end-user correctly, in the use of the software. Thus, not only more people could understand the development of a given solution, but even more, people would be able to use it in their researches, increasing the number of citations of the original work and leading to a wider spread of the solution. Adoption of a software development methodology could definitively solve this problem, but it would demand deep knowledge of one of these methodologies by Bioinformatics researchers, which usually is not the case \cite{verma2013lack}. As a way of simplifying the adoption of good Software Engineering practices without formal training, one could adopt recommendations like \citet{karimzadeh2018top}'s, focusing in the end-user documentation.

In this study, we analyzed Biopipeline \cite{de2011genome}, a bioinformatics software used to predict miRNAs within a genome, starting from the source code and building documentation, for both maintenance and the end-user. We followed \citet{karimzadeh2018top} recomendations where applicable, and formatted the end user documentation as simple as possible.

\section{Materials and methods}\label{sec2}

\subsection{The Biopipeline software}\label{subsec2-1}

The analyzed software is named Biopipeline and was written to identify conserved miRNAs within a genome. The source code of the Biopipeline software was kindly provided by Laboratório de Informática e Análises Moleculares (LBAM) of the Federal University of Uberlândia, at Patos de Minas. It was imported in Atom text editor, using the Perl extension. Preliminary analysis showed that it was written in Perl using the Structured Programming paradigm. It is comprised of a series of procedures called functions and numbered from F1 to F50, each one of them performing a specific task.

\subsection{Technical documentation}\label{subsec2-2}

For a better understanding of the functionalities, and to guide the development of the end-user documentation, we built a block diagram using Dia 0.97, and a two-level data flow diagram using LibreOffice Draw. We made these documents available so that they can also be useful for software maintenance.

\subsection{Guidelines for end user documentation}\label{subsec2-3}

In order to provide a simpler way for researchers to document software already implemented, we followed the considerations for the creation of bioinformatics software documentation proposed by \citet{karimzadeh2018top}. Their work provide various considerations about bioinformatics software documentation, being the minimum set of documentation comprised of a GitHub or Bitbucket page with the source code and an issue tracker; a \textit{Readme} file containing an installation guide, a quick start guide, and the file formats used for inputs and outputs; and a reference manual with a detailed description of each user-configurable parameter. Since the authors of Biopipeline are not releasing the source code as open-source at the moment, we have launched a GitHub page (available at https://anonymous.4open.science/r/biopipeline-docs-110C) only as a documentation repository, not providing source code or issue tracker. Additionally, as the authors have stated that there are no user-configurable parameters, and all parameters are set within the source code with optimized values, we also have not written a reference manual.

\section{Results}\label{sec3}

\subsection{Source code analysis and technical documentation}\label{subsec3-1}

Analysis of the source code revealed that the software was written in Perl using the Structured Programming paradigm. It is comprised of a series of procedures called functions and numbered from F1 to F50, each one of them performing a specific task. In addition, no Object Oriented Programming was identified, like classes or instances.

After realizing what methodology was used when programming, we identified that aside from the processing conducted by original code, additional, external tools are used to perform important parts of the overall processing, using a concept called pipeline \cite{kumar2007bioinformatics}, which means that various steps of processing are executed in a specific sequence.

This so-called “reverse engineering” made it possible to understand exactly how the software does its work. The first input of the system is a fasta file containing a complete genome. This file is conducted to a preliminary analysis using \textit{einverted EMBOSS} and \textit{BLASTn} software, configured with optimized parameters identified by the authors. This step returns as output the structures identified as hairpins and homolog pre-miRNAs from miRBase. Those sequences are archived in various files for parallel processing.

Those files are then processed by \textit{RNAfold}, then \textit{BLASTn} again, to identify similarities between the found sequences and known miRNAs. Next, the sequences that resemble coding genes, non-coding RNA (ncRNA), and repeating sequences are discarded. Then, another selection is performed by \textit{BLASTn} with more specific parameters.
After all these steps are executed, the files with the sequences are joined, resulting in a single fasta file containing all miRNAs found. The overall processing is represented in Figure 1, by means of a block diagram which was divided into three parts for better visualization.

\begin{figure*}[!t]%
\centering
\includegraphics[width=0.65\textwidth]{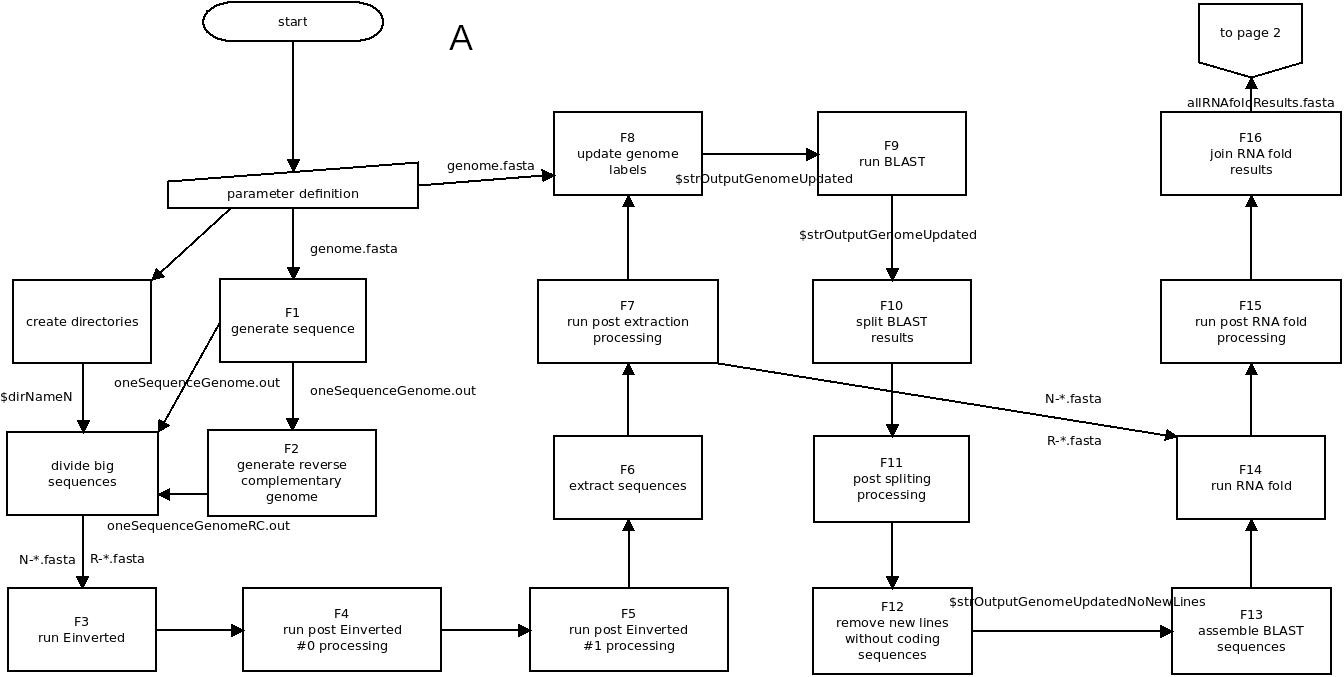}
\label{fig1a}
\end{figure*}

\begin{figure*}[!t]%
\centering
\includegraphics[width=0.65\textwidth]{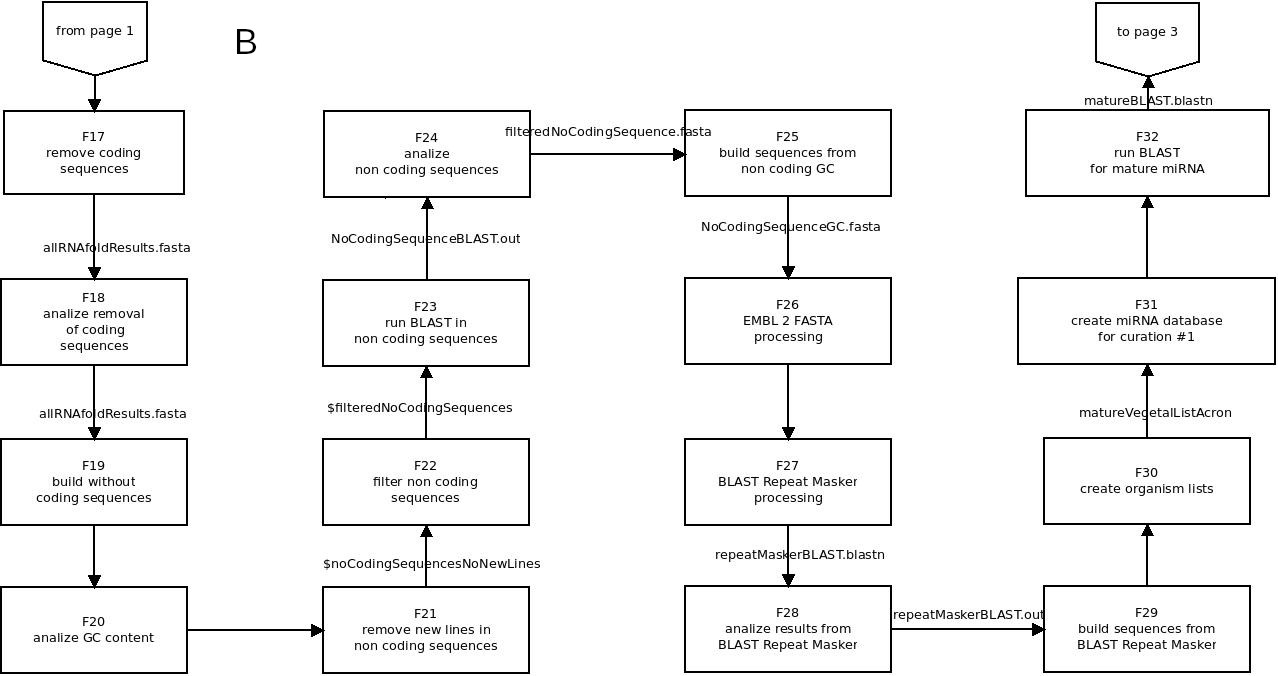}
\label{fig1b}
\end{figure*}

\begin{figure*}[!t]%
\centering
\includegraphics[width=0.65\textwidth]{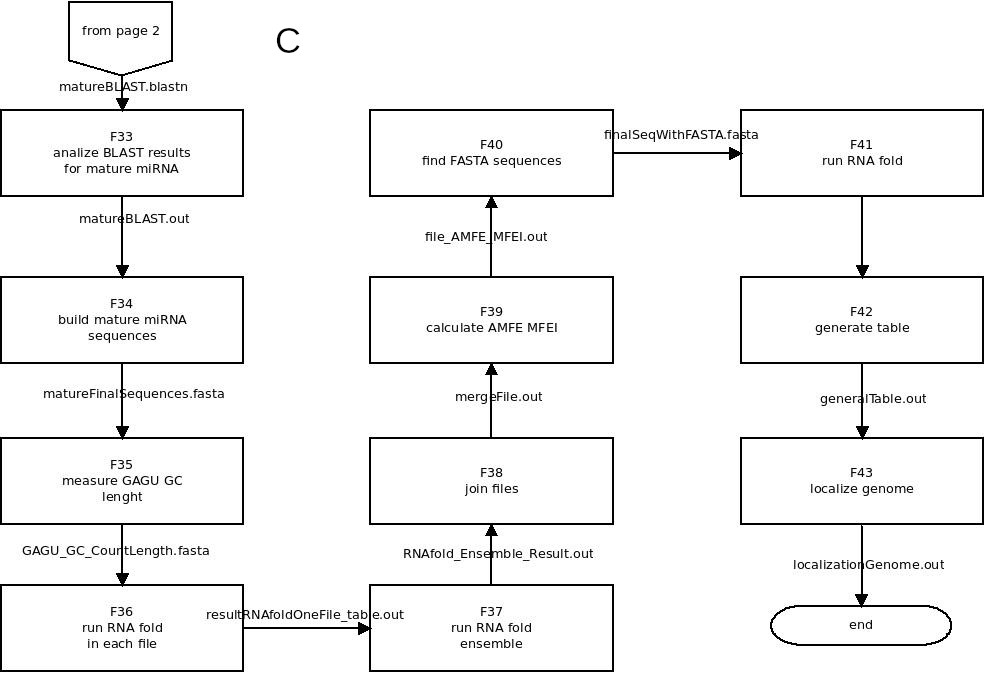}
\caption{Block diagram showing an overview of the system, with inputs and outputs. It was divided into three parts (A)(B)(C) for better visualization. Each “FXX” represents a function defined in the Biopipeline algorithm, and it’s possible to see that some of them are performed by external software.
}\label{fig1c}
\end{figure*}

The block diagram is useful to translate the source code to a visual language, but showed itself very large and complex, and not the best representation of the system, serving instead as a starting point for the understanding of the functionalities and how Biopipeline uses external software to perform specific tasks.

\begin{figure*}[!t]%
\centering
\includegraphics[width=0.40\textwidth]{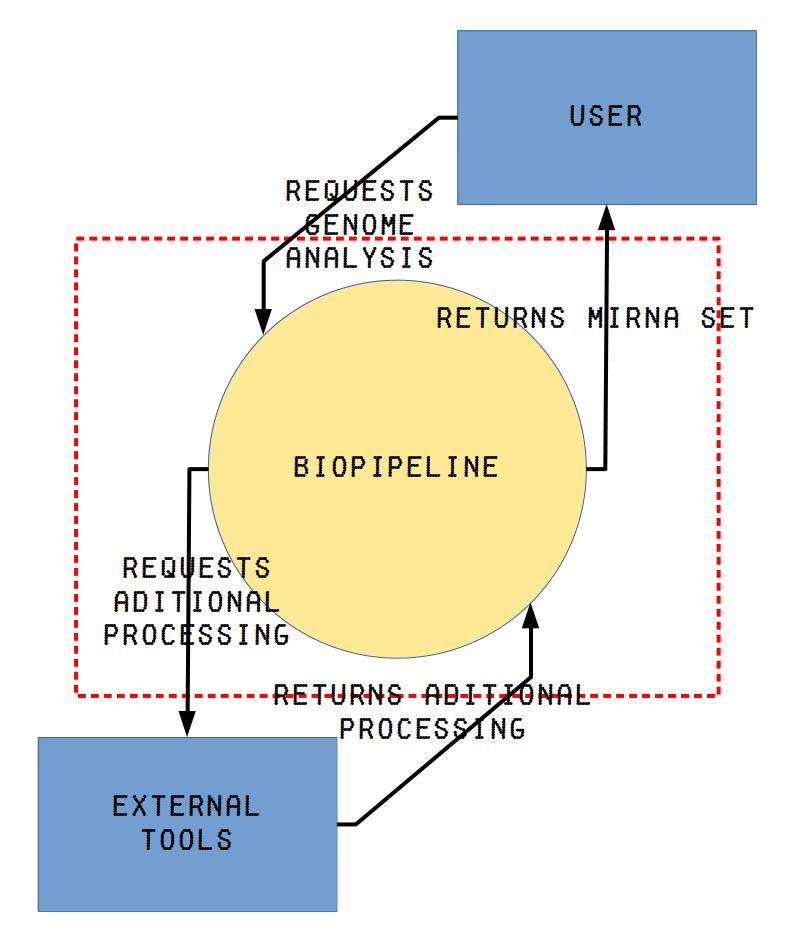}
\caption{Level 0 Data Flow Diagram, showing an overview of the system and the existence of additional processing done by external tools
}\label{fig4}
\end{figure*}

A data flow diagram is a better, clearer representation of the system operation. It shows the relationships between Biopipeline and its external dependencies and the data flow in a more direct way. We built it as a two-level diagram. Level 0, showed in Figure \ref{fig4} provides an overview of what the software does and makes clear that it uses external tools for additional processing. It also shows the inputs and outputs in a simplified way, making it clear that the main input is the genome, which is processed by Biopipeline and external software, and the output is the set of miRNAs found.
	
\begin{figure*}[!t]%
\centering
\includegraphics[width=0.80\textwidth]{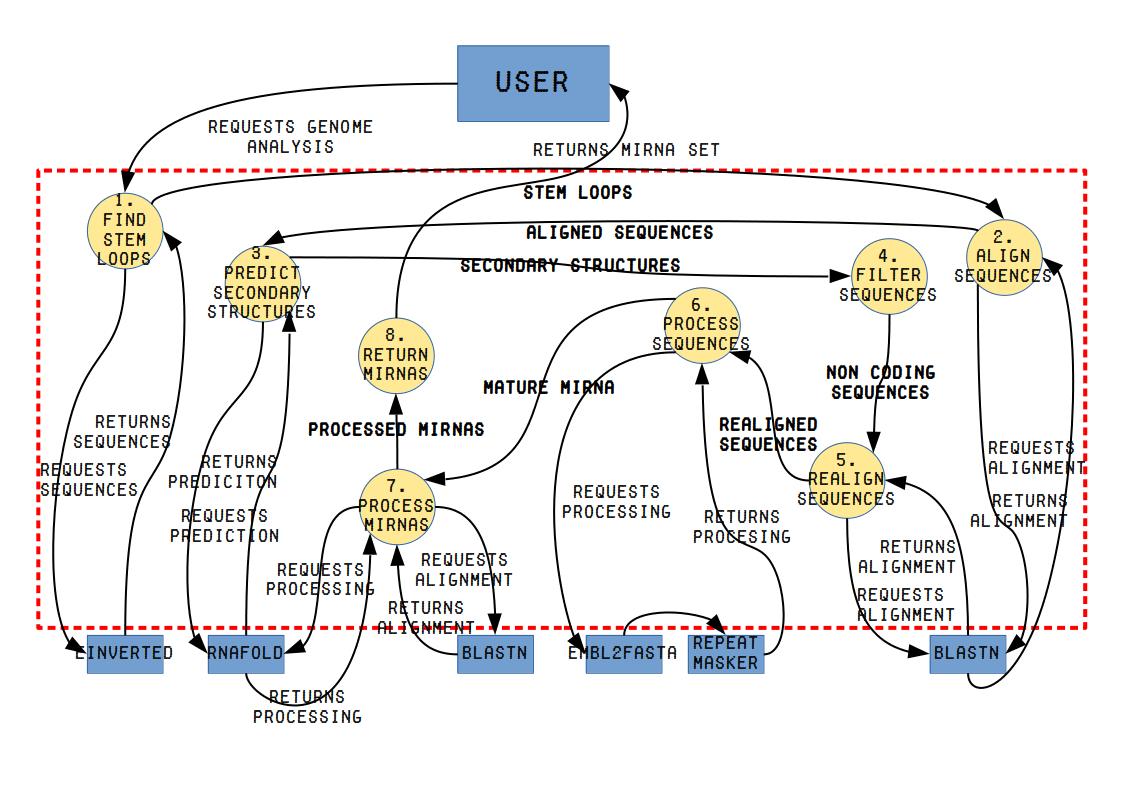}
\caption{Level 1 Data Flow Diagram showing details of the processing performed on the file containing the genome, including what and when external tools (\textit{einverted}, \textit{RNAFold}, \textit{Blastn}, \textit{Embl2Fasta}, and \textit{Repeat Masker}) are called. One of these tools, \textit{Blastn}, is shown twice for aesthetic reasons
}\label{fig5}
\end{figure*}

Nonetheless, this level of detail is insufficient to show how the software analysis the genome until finding the miRNAs. In order to reach this detailing, a second level of the data flow diagram is presented in Figure \ref{fig5}. This Level 1 diagram shows what kind of processing is performed on the initial genome file, and exactly when external tools are called. There is no direct reference to the actual functions in the algorithm, contrary to the block diagram.
	
\subsection{End-user documentation}\label{subsec3-2}

The technical documentation above was produced to allow a better understanding of the software. Both block diagram and data flow diagram were useful to retrieve important information used in the production of the end-user documentation and are also useful for maintenance tasks.

In order to produce relevant yet understandable documentation, we followed \citet{karimzadeh2018top} recommendations where applicable. This led us to build a \textit{Readme} file containing an overview, installation guide, and user guide. Nonetheless, we have opted to launch a GitHub/Bitbucket page only as a documentation repository, and not as recommended, because the authors are not releasing the source code as open-source at the moment; and a reference manual is not provided because all parameters were optimized and set in the source code by the authors and are not user-configurable.

\begin{figure*}[!t]%
\centering
\includegraphics[width=0.80\textwidth]{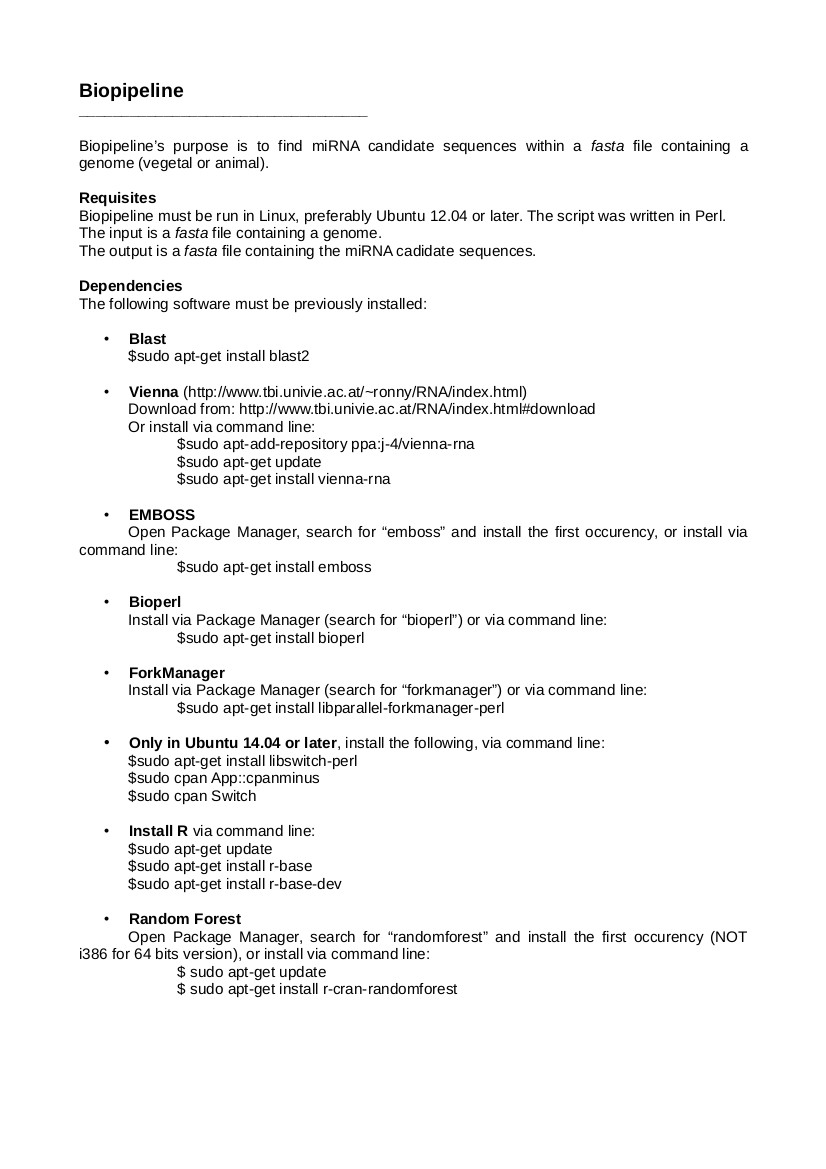}
\caption{The end-user documentation includes an overview of the software along with a list of dependencies that must be satisfied before installing it. The document was formatted in a simple readable way 
}\label{fig6}
\end{figure*}

\begin{figure*}[!t]%
\centering
\includegraphics[width=0.80\textwidth]{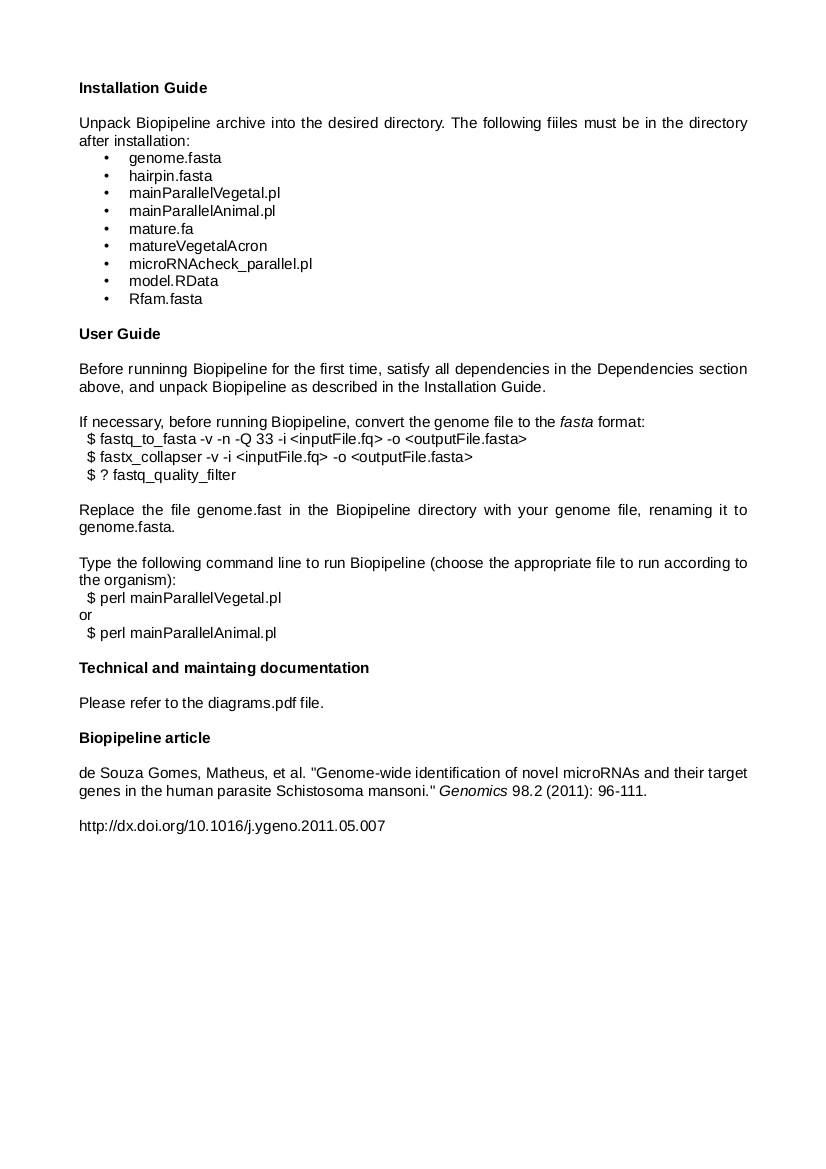}
\caption{The installation guide contains straight forward instructions. The user guide shows command line examples and instructions on the input format. There is a reference to another document containing the technical documentation and a reference to the original Biopipeline manuscript 
}\label{fig7}
\end{figure*}

The \textit{Readme} file was formatted as simple as possible, yet containing all information needed in order to solve dependencies, as shown in Figure \ref{fig6}, install and run the software, as shown in Figure \ref{fig7}. It provides the minimum requirements of the system, examples of command lines, and references the technical documentation and the original Biopipeline article.

%\FloatBarrier

\section{Discussion}\label{sec4}

Documenting software can be a tedious task when one is anxious to finish development, but it is a task better done during this period. However, reverse engineering of a piece of software can lead to the building of good documentation, both maintaining and end-user documentation. Many authors have demonstrated the importance of documenting bioinformatics software, such as \citet{leprevost2014best}, \citet{karimzadeh2018top}, \citet{kumar2007bioinformatics}. 

Amongst the reasons to invest time and efforts in the documentation task are: improvement of software usability, continuity of the software development, reproducibility of experiments, better software maintenance, reducing of support requests for the authors, and increasing of citations of the original work. Thus, documenting bioinformatics software can benefit the authors, researchers, scientific community, and general community.

In our efforts to document Biopipeline, we showed that is possible to write useful documentation for already implemented software. We also showed that this can be done by means of reverse engineering, and maybe done on software written by others. Additionally, we showed that the documentation can (and sometimes must) be simple, understandable, and yet bring useful information. These results have the potential to help encourage researchers to document their software (and software developed by other researchers), thus benefiting the community.




%\bibliographystyle{IEEEtran}
\bibliography{main}

\begin{thebibliography}{11}
\providecommand{\natexlab}[1]{#1}
\providecommand{\url}[1]{\texttt{#1}}
\expandafter\ifx\csname urlstyle\endcsname\relax
  \providecommand{\doi}[1]{doi: #1}\else
  \providecommand{\doi}{doi: \begingroup \urlstyle{rm}\Url}\fi

\bibitem[de~Souza~Gomes et~al.(2011)de~Souza~Gomes, Muniyappa, Carvalho,
  Guerra-S{\'a}, and Spillane]{de2011genome}
M.~de~Souza~Gomes, M.~K. Muniyappa, S.~G. Carvalho, R.~Guerra-S{\'a}, and
  C.~Spillane.
\newblock Genome-wide identification of novel micrornas and their target genes
  in the human parasite schistosoma mansoni.
\newblock \emph{Genomics}, 98\penalty0 (2):\penalty0 96--111, 2011.

\bibitem[Jawdat(2006)]{jawdat2006era}
D.~Jawdat.
\newblock The era of bioinformatics.
\newblock In \emph{2006 2nd International Conference on Information \&
  Communication Technologies}, volume~1, pages 1860--1865. IEEE, 2006.

\bibitem[Karimzadeh and Hoffman(2018)]{karimzadeh2018top}
M.~Karimzadeh and M.~M. Hoffman.
\newblock Top considerations for creating bioinformatics software
  documentation.
\newblock \emph{Briefings in Bioinformatics}, 19\penalty0 (4):\penalty0
  693--699, 2018.

\bibitem[Kipyegen and Korir(2013)]{kipyegen2013importance}
N.~J. Kipyegen and W.~P. Korir.
\newblock Importance of software documentation.
\newblock \emph{International Journal of Computer Science Issues (IJCSI)},
  10\penalty0 (5):\penalty0 223, 2013.

\bibitem[Kumar and Chordia(2017)]{kumar2017role}
A.~Kumar and N.~Chordia.
\newblock Role of bioinformatics in biotechnology.
\newblock \emph{Res Rev Biosci}, 12\penalty0 (1):\penalty0 116, 2017.

\bibitem[Kumar and Dudley(2007)]{kumar2007bioinformatics}
S.~Kumar and J.~Dudley.
\newblock Bioinformatics software for biologists in the genomics era.
\newblock \emph{Bioinformatics}, 23\penalty0 (14):\penalty0 1713--1717, 2007.

\bibitem[Lawlor and Walsh(2015)]{lawlor2015engineering}
B.~Lawlor and P.~Walsh.
\newblock Engineering bioinformatics: building reliability, performance and
  productivity into bioinformatics software.
\newblock \emph{Bioengineered}, 6\penalty0 (4):\penalty0 193--203, 2015.

\bibitem[Leprevost et~al.(2014)Leprevost, Barbosa, Francisco, Perez-Riverol,
  and Carvalho]{leprevost2014best}
F.~d.~V. Leprevost, V.~C. Barbosa, E.~L. Francisco, Y.~Perez-Riverol, and P.~C.
  Carvalho.
\newblock On best practices in the development of bioinformatics software.
\newblock \emph{Frontiers in genetics}, 5:\penalty0 199, 2014.

\bibitem[Lethbridge et~al.(2003)Lethbridge, Singer, and
  Forward]{lethbridge2003software}
T.~C. Lethbridge, J.~Singer, and A.~Forward.
\newblock How software engineers use documentation: The state of the practice.
\newblock \emph{IEEE software}, 20\penalty0 (6):\penalty0 35--39, 2003.

\bibitem[Parnas(2011)]{parnas2011precise}
D.~L. Parnas.
\newblock Precise documentation: The key to better software.
\newblock In \emph{The Future of Software Engineering}, pages 125--148.
  Springer, 2011.

\bibitem[Verma et~al.(2013)Verma, Gesell, Siy, and Zand]{verma2013lack}
D.~Verma, J.~Gesell, H.~Siy, and M.~Zand.
\newblock Lack of software engineering practices in the development of
  bioinformatics software.
\newblock \emph{ICCGI}, 2013:\penalty0 57--62, 2013.

\end{thebibliography}

\end{document}